\documentclass[aps,pra,preprint,groupedaddress,showpacs,%
tightenlines,
nofootinbib,floatfix]{revtex4}

\usepackage{amsmath,amssymb,amsfonts,bm}
\usepackage{graphicx}
\usepackage{mathrsfs} 
\usepackage{hyperref}
\usepackage{txfonts}

\renewcommand{\vec}[1]{\bm{\mathrm{#1}}} %

\begin{document}

\title{Cold neutron scattering in imperfect deuterium crystals} 

\author{Andrzej Adamczak}
\email{andrzej.adamczak@ifj.edu.pl}
\affiliation{Institute of Nuclear Physics, Polish Academy of Sciences, 
  Radzikowskiego~152, PL-31342~Krak\'ow, Poland}

\date{\today}

\begin{abstract}
  The differential cross sections for cold neutron scattering in mosaic
  deuterium crystals have been calculated for various target
  temperatures.  The theoretical results are compared with the recent
  experimental data for the neutron wavelengths
  $\lambda\approx$~1--9~\AA. It is shown that the structures of observed
  Bragg peaks can be explained by the mosaic spread of about $3^{\circ}$
  and contributions from a~limited number of crystal orientations. 
  Such a~crystal structure should be also taken into account in
  ultracold neutron upscattering due to the coherent phonon annihilation
  in solid deuterium.  
\end{abstract}


\maketitle


\section{Introduction}
\label{sec:intro}

The aim of this work is a~calculation of the differential cross sections
for cold neutron scattering in solid molecular deuterium (sD$_2$) at
various temperatures and comparison with the recent experimental
data~\cite{kasp08,atch09}. Deceleration of cold neutrons in sD$_2$
converters is an effective method for production of high-intensity
ultracold neutron
sources~\cite{golu83,sere94,sere00,soun04,bode04,atch05,frei07,kasp08,atch09}.
Therefore knowledge of the neutron cross sections is indispensable for
effective projecting such sources. Ultracold neutrons (UCN) are applied
for studies of fundamental properties of the
neutron~\cite{sere05,bake06}. Also, they can be used for surface
physics~\cite{golu91}, investigations of the neutron mirror-neutron
oscillations~\cite{ban07} and neutron quantum states in the
gravitational potential~\cite{nesv03}.

The experimental total cross sections~\cite{kasp08,atch09} in the
Bragg-scattering region display a~certain number of pronounced peaks.
Their heights decrease when sD$_2$ temperature is rising from 8~K to
18~K. However, for a fixed crystal, the structure of Bragg peaks does
not practically change with varying temperature. It is well known that
sD$_2$ at such temperatures and under low pressure has the hcp
structure~\cite{silv80,soue86}, which has been confirmed using the Raman
spectroscopy~\cite{bode04}. A~simple calculation of the Bragg cross
sections presented in Ref.~\cite{kasp08} does not lead to agreement with
the experiment, both for the random polycrystalline hcp structure and
for a~few specifically oriented hcp polycrystals. The magnitudes of the
theoretical peaks are too large. Also, these peaks are sharp, while the
experimental peaks are quite broad. Moreover, the observed and
calculated locations of the Bragg peaks do not fully coincide. A more
advanced calculation of the total cross section for a random
polycrystalline sD$_2$~\cite{gran09}, with incoherent processes taken
into account, also disagrees with the experiment in the Bragg-scattering
region.

A calculation of the cross sections presented in this paper includes
both the elastic and inelastic processes in the sD$_2$ lattice and in
D$_2$ molecules. Rotational and vibrational transitions in these
molecules, which are due to collisions with neutrons, are taken into
account. Interactions of neutrons with phonons are described in the
incoherent approximation while the incoherent and coherent elastic cross
sections are treated separately.  The coherent elastic scattering of
neutrons is considered both for the random polycrystalline hcp structure
and for different orientations of the hcp monocrystal.

The calculated total cross sections are compared with the available data
for the neutron wavelengths
$\lambda\approx$~1--9~\AA~\cite{kasp08,atch09}, with the experimental
uncertainty of neutron wavelength taken into account. Since this
uncertainty is much smaller than the observed widths of the Bragg
peaks~\cite{kasp_priv}, an explanation of this effect by the presence of
mosaic imperfections in the sD$_2$ crystals is studied. Most crystals
are perfect only in very small domains (mosaic blocks) on the order of
1000~\AA, which are separated by displacements and distortions in the
lattice. These domains are almost parallel, with small deviations of
their orientations from the main direction. Such angular deviations are
usually well-described by a Gaussian distribution~\cite{zach45}. The
mosaic spread may reach even a~few degrees in certain crystals.

Another experimental problem is connected with different structures of
the sD$_2$ crystals, which are grown even in very similar conditions. As
a result, the experimental Bragg cross sections for the sD$_2$ crystals,
obtained by freezing D$_2$ gas and two samples of liquid D$_2$, are
different~\cite{kasp08,atch09}. In this paper, such a phenomenon is
ascribed to different orientations of large monocrystals (much greater
than the mosaic blocs) that are present in a given sD$_2$ target. These
orientations differ much greater than those of microscopic mosaic blocs
within a single monocrystal. Such a model of a real deuterium target is
suggested by the optical observations of growing
sD$_2$~\cite{kasp08,bode04}.

\section{Method of calculation}
\label{sec:method}

Interaction of the neutron with a single deuterium nucleus is described
using the Fermi pseudopotential~\cite{ferm36}. The partial differential
cross section for neutron scattering in sD$_2$ can be expressed in terms
of the Van Hove response function~$\mathcal{S}$~\cite{vanh54}, which
depends on the target properties for fixed momentum transfer and energy
transfers. When the impinging neutron causes a rotational-vibrational
transition in a deuterium molecule bound in the sD$_2$ lattice, the
corresponding cross section can be expressed as follows
\begin{equation}
  \label{eq:x_rovib}
  \left(\frac{\partial^2\sigma}
    {\partial\Omega\partial\varepsilon'}\right)_{0n} = 
  \frac{k'}{k}\, \sigma_{0n} \,\mathcal{S}_i(\vec{\kappa},\omega) \,, 
\end{equation}%
where $\mathcal{S}_i$ denotes the incoherent response function, which is
a fraction of the total response function $\mathcal{S}$~\cite{love84}.
The energy transfer~$\omega$ and the momentum transfer~$\vec\kappa$ to
the sD$_2$ lattice are defined as (in the atomic units)
\begin{equation}
  \label{eq:transol_def}
  \omega = \varepsilon -\varepsilon' -\Delta E \,, \qquad\qquad
  \vec\kappa = \vec{k} - \vec{k}' ,
\end{equation}%
where $\varepsilon$ and $\varepsilon'$ stand for the initial and final
kinetic energies of the neutron and $\Delta E$ is the internal-energy
change of the target molecule due to the rotational- vibrational
transition. Vectors $\vec{k}$ and~$\vec{k}'$ denote the initial and
final neutron momenta, respectively. The usual relations between the
kinetic energies $\varepsilon$ and $\varepsilon'$ and the corresponding
absolute values $k$ and $k'$ of neutron momenta are fulfilled
\begin{equation}
  \label{eq:en_mom_rel}
  \varepsilon  = k^2/2m_n \,,  \qquad  \varepsilon' = k'^2/2m_n \,,
\end{equation}
$m_n$ being the neutron mass. Function $\sigma_{0n}$ in
Eq.~(\ref{eq:x_rovib}) is the squared modulus of neutron scattering
amplitude for a single bound molecule
\begin{equation}
  \label{eq:sig_0n}
  \sigma_{0n} = \left|\mathcal{F}_{0n}\right|^2 ,
\end{equation}
averaged over the total spin of the neutron and D$_2$ system. The
indices $(0)=(0K)$ and $(n)=(\nu{}K')$ denote the initial and final
rotational-vibrational states, respectively. In the case of considered
target temperatures, all the molecules are initially in the ground
vibrational state. The final vibrational state is characterized by the
quantum number~$\nu$. The quantum numbers $K$ and $K'$ correspond to the
initial and final rotational states of D$_2$.

When $\Delta{}E=0$, for all the molecules in sD$_2$, neutron scattering
may be strictly elastic (no simultaneous phonon processes) or
quasielastic (with phonon creation or annihilation). It is conventional
to separately calculate the coherent and incoherent fractions of the
corresponding differential cross section. The coherent cross section,
which displays strong interference effects from all lattice sites, takes
the general form~\cite{love84}
\begin{equation}
  \label{eq:xcoh_def}
  \left(\frac{\partial^2\sigma}
    {\partial\Omega\partial\varepsilon'}\right)_\mathrm{coh} = 
    \frac{k'}{k}\, \sigma_\mathrm{coh}\,
    \mathcal{S}(\vec{\kappa},\omega),  
\end{equation}
per one D$_2$ molecule. The coherent elastic cross section
$\sigma_\mathrm{coh}$ for a~single bound molecule is defined as
\begin{equation}
  \label{eq:amp_coh}
  \sigma_\mathrm{coh} = \bigl|\overline{\mathcal{F}_{00}} \bigr|^2 .   
\end{equation}%
The horizontal bar denotes here averaging over a~random distribution of
the total spin of the $n+\mathrm{D}_2$ system and over the molecular
rotational population in the lattice. The value~$I$ of total spin of the
symmetric molecule D$_2$ is equal to $I=0$, 2 for even rotational
numbers~$K$ (ortho-D$_2$) and $I=1$ for odd rotational numbers
(para-D$_2$). In solid deuterium, the two rotational states $K=0$ and
$K=1$ are usually present, since the rotational relaxation
$K=1\to{}K'=0$ is very slow without a catalyst~\cite{soue86}. It is
assumed that there is no correlation between the spin~$\vec{I}$ and
the location of D$_2$ in the sD$_2$ lattice.

The incoherent cross section for neutron scattering in sD$_2$
($\Delta{}E=0$) is expressed by the formula similar to 
Eq.~(\ref{eq:x_rovib})~\cite{love84}
\begin{equation}
  \label{eq:xinc_def}
  \left(\frac{\partial^2\sigma}
    {\partial\Omega\partial\varepsilon'}\right)_\text{inc} = 
  \frac{k'}{k}\, \sigma_\mathrm{inc}\, 
  \mathcal{S}_i(\vec{\kappa},\omega) , 
\end{equation}%
where $\sigma_\mathrm{inc}$ is the incoherent elastic cross section for
neutron scattering from a~single bound molecule
\begin{equation}
  \label{eq:amp_incoh}
  \sigma_\mathrm{inc} = \overline{\left|\mathcal{F}_{00}\right|^2}
  -\bigl| \overline{\mathcal{F}_{00}}\bigr|^2 =
  \overline{\left|\mathcal{F}_{00}\right|^2}-\sigma_\mathrm{coh} \,.
\end{equation}%

The functions $\sigma_{0n}$, $\sigma_\mathrm{coh}$, and
$\sigma_\mathrm{inc}$ have been calculated similarly as the analogous
quantities for the muonic atom scattering in gaseous and solid
hydrogenic targets~\cite{adam06,adam07}. Since the muonic atoms are
small neutral objects, the methods of evaluation of the cross sections
for these atoms are almost the same as for neutrons. Therefore, it is
sufficient to use the neutron mass, scattering lengths, and appropriate
spin correlations~\cite{hame46,youn64} in the general equations from
Refs.~\cite{adam06,adam07}. As a result, in the case of $n+\mathrm{D}_2$
scattering one obtains
\begin{equation}
  \label{eq:sigmol_inc_inel}
  \sigma_\mathrm{0n} =
  \begin{cases}
    (4b_\mathrm{coh}^2+\tfrac{5}{2} b_\mathrm{inc}^2)\, f_{KK'}(\kappa)
    & \mathrm{~even~} K \mathrm{~and~} K' , \\[4pt]
    \tfrac{3}{2}\, b_\mathrm{inc}^2\, f_{KK'}(\kappa)
    & \mathrm{~even~} K, \mathrm{~odd~} K' , \\[4pt]
    3b_\mathrm{inc}^2\, f_{KK'}(\kappa) 
    & \mathrm{~odd~} K, \mathrm{~even~} K' , \\[4pt]
    (4b_\mathrm{coh}^2+b_\mathrm{inc}^2) f_{KK'}(\kappa)
    & \mathrm{~odd~} K \mathrm{~and~} K' ,
  \end{cases}
\end{equation}%
where $b_\mathrm{coh}=6.671$~fm and $b_\mathrm{inc}=4.04$~fm denote,
respectively, the coherent and incoherent scattering length for neutron
scattering from a~bound deuterium nucleus. The molecular form factor
$f_{KK'}$ is defined as
\begin{equation}
  \label{eq:amp_squared}
  f_{KK'}(\kappa) = \sum_l \mathcal{W}_{K'lK} 
  |\mathcal{D}_{\nu l}(\kappa)|^2 ,
\end{equation}
with the angular-momentum factor $\mathcal{W}_{K'lK}$ 
\begin{equation}
  \label{eq:Wig_K'lK}
  \mathcal{W}_{K'lK} \equiv (2K'+1)(2l+1)
      \begin{pmatrix}
      K' & l & K \\
      0  & 0 & 0
    \end{pmatrix}^2 ,  \quad
     \mathcal{W}_{K0K} = 1 
\end{equation}
expressed by the Wigner $3j$~symbols. Equation~(\ref{eq:Wig_K'lK}) has
been obtained upon averaging $\sigma_\mathrm{0n}$ over projections of
the initial angular momentum of D$_2$ and summing over projections of
the final angular momentum of this molecule.  Function
$\mathcal{D}_{\nu{}l}$ in Eq.~(\ref{eq:amp_squared}) denotes the radial
matrix element
\begin{equation}
  \label{eq:molfac}
  \begin{split}
    \mathcal{D}_{\nu l}(\kappa) & \equiv \int_0^{\infty} \mathrm{d}R 
    \, \chi_\nu(R) \mathrm{j}_l(\kappa R/2) \chi_0(R) \,, \\[4pt]
    \mathcal{D}_{\nu l}(\kappa) & \xrightarrow[\kappa\to\, 0]{}  
    \begin{cases}
      1&  \mathrm{~when~} \nu=0 \mathrm{~and~} l=0 \,, \\
      0&  \mathrm{~otherwise,}
    \end{cases}
  \end{split}
\end{equation}
in which $\mathrm{j}_l$ is the $l$th spherical Bessel function. The
radial wave function of the D$_2$ molecule for the vibrational
state~$\nu$ is denoted here by~$\chi_\nu$ and $R$ stands for the
internuclear distance in~D$_2$. Numerical evaluation of the matrix
elements (\ref{eq:molfac}) has been performed assuming that the
vibrations of D$_2$ are harmonic and that there is no coupling between
the molecular vibrations and rotations. Employing the harmonic-potential
model of D$_2$ instead of the rigid-rotor model leads to accurate cross
sections even at high collision energies.

The coherent elastic cross section~(\ref{eq:amp_coh}) is expressed by
the coherent scattering length~$b_\mathrm{coh}$ and
the radial matrix element~(\ref{eq:molfac}) for $\nu=0$ and $l=0$ 
\begin{equation}
  \label{eq:sigmol_coh}
  \sigma_\mathrm{coh} =
  \begin{cases}
    4 b_\mathrm{coh}^2 |\mathcal{D}_{00}(\kappa)|^2 
    & \mathrm{~for~} K = K' , \\[2pt]
    0 
    & \mathrm{~otherwise}.
  \end{cases}
\end{equation}%

In the case of harmonic hcp lattice, the incoherent response function
$\mathcal{S}_i$ in Eqs.~(\ref{eq:x_rovib}) and~(\ref{eq:xinc_def}) is
approximated by the phonon expansion of $\mathcal{S}_i$ for a Bravais
cubic lattice~\cite{sing60,love84}
\begin{equation}
  \label{eq:Sinc_phon_exp}
  \mathcal{S}_i(\vec{\kappa},\omega) = \exp(-2W) 
     \left[ \delta(\omega) +\sum_{n=1}^{\infty} g_n(\omega) 
     \, \frac{(2W)^n}{n !} \right],
\end{equation}
in which $\exp(-2W)=\exp[-2W(\vec\kappa)]$ denotes the Debye-Waller
factor. The functions $g_n$ are defined as
\begin{equation}
  \label{eq:g_n}
  \begin{split}
    g_1(\omega) = & \frac{1}{2W}\,\frac{\kappa^2}{2M_\mathrm{mol}} 
    \frac{Z(\omega)}{\omega} \left[\,n_\mathrm{B}(\omega)+1\right] , 
    \\[3pt]
    g_n(\omega) = & \int_{-\infty}^{\infty} \mathrm{d} \omega' \,
    g_1(\omega-\omega')\, g_{n-1}(\omega') \,, \\[3pt]
    & \int_{-\infty}^{\infty}\mathrm{d} \omega \, g_n(\omega)  = 1 \,,
  \end{split}
\end{equation}
where $M_\mathrm{mol}$ is the mass of D$_2$ molecule, $Z(\omega)$ is the
normalized density of vibrational states of the lattice, and
$n_\mathrm{B}$ denotes the Bose phonon-population factor
\begin{equation}
  \label{eq:Bose_fac}   
  n_\mathrm{B}(\omega)=\left[\,\exp(\omega/k_\mathrm{B}T)-1\right]^{-1},
\end{equation}%
for a~given temperature~$T$ ($k_\mathrm{B}$ is the Boltzmann constant).
The first term in the phonon expansion (\ref{eq:Sinc_phon_exp})
corresponds to recoil-less scattering from a rigid sD$_2$ lattice.
When $\Delta{}E=0$, such a scattering cannot change the neutron kinetic
energy because of a very large mass of the sD$_2$ target. The next terms
describe neutron scattering and possible rotational-vibrational
transitions in D$_2$ with simultaneous creation ($\omega>0$) or
annihilation ($\omega<0$) of phonons. In particular, the second term in
Eq.~(\ref{eq:Sinc_phon_exp}) is connected with one-phonon
processes. Equations~(\ref{eq:g_n}) have been derived assuming that 
$Z(-\omega)\equiv{}Z(\omega)$.

Evaluation of the incoherent cross section has been performed using the
isotropic Debye model, characterized  by the following density of
vibrational states
\begin{equation}
  \label{eq:Z_Debye}
  Z(\omega) =
  \begin{cases}
    3\, \omega^2/\omega_\mathrm{D}^3 & \mathrm{~if~} \,
    \omega \leq \omega_\mathrm{D} \\
    0 & \mathrm{~if~} \, \omega > \omega_\mathrm{D} \,,
  \end{cases}
\end{equation}%
where $\omega_\mathrm{D}=k_\mathrm{B}\varTheta_\mathrm{D}$ denotes the
Debye energy corresponding to the Debye temperature
$\varTheta_\mathrm{D}$. In the case of solid hydrogenic targets at low
pressures, $\varTheta_\mathrm{D}$ equals about 100~K and it slowly
decreases with rising target temperature~\cite{silv80}. For the
calculation of cross sections at various temperatures, the following
estimation of the Debye temperature
\begin{equation}
  \label{eq:edeb-sound}
  \varTheta_\text{D} =  2^{1/2}(3\pi^2)^{1/3} \beta\,
  \frac{c_\mathrm{s}}{a_\mathrm{hcp}\, k_\mathrm{B}} \,,
\end{equation}
has been used. The sound velocity $c_\mathrm{s}$ and the molar volume of
sD$_2$ for calculation of the lattice constant~$a_\mathrm{hcp}$ have
been taken from Ref.~\cite{soue86}. A constant correction factor
$\beta=1.085$ in Eq.~(\ref{eq:edeb-sound}) is introduced to fit the
experimental value of 114~K for sD$_2$ at zero
temperature~\cite{niel71}. As a result, one obtains
$\varTheta_\text{D}=110$~K for 8-K sD$_2$ and $\varTheta_\text{D}=104$~K
for 18-K sD$_2$.

The exponent $2W$ of the Debye-Waller factor for a cubic lattice takes
the form~\cite{love84}
\begin{equation}
  \label{eq:2W_u}
  2W =\tfrac{1}{3}\langle \vec{u}^2 \rangle \, \kappa^2 \,,
\end{equation}
where $\langle\vec{u}^2\rangle$ is the mean square displacement of the
molecule from its lattice site. In the case of harmonic crystal,
Eq.~(\ref{eq:2W_u}) can be estimated using the expression
\begin{equation}
  \label{eq:2W}
  2W = \frac{\kappa^2}{2M_\mathrm{mol}} \int_0^{\infty} 
  \mathrm{d}\omega \, \coth\left(\frac{\omega}{2k_\mathrm{B}T}\right)
  \frac{Z(\omega)}{\omega} \,.
\end{equation}
This expression is employed here as a good approximation for a harmonic
hcp lattice. For a 5-K sD$_2$ crystal, Eqs. (\ref{eq:edeb-sound}) and
(\ref{eq:2W}) lead to the root-mean-square displacement of 0.50~\AA,
which is in excellent agreement with the experimental result from
Ref.~\cite{niel71}.

The calculation of all phonon cross sections is performed here in the
incoherent approach, which is achieved by substituting
$\sigma_\mathrm{inc}+\sigma_\mathrm{coh}$ for $\sigma_\mathrm{inc}$ in
the nonelastic fraction of Eq.~(\ref{eq:xinc_def}). As a result, only
the elastic Bragg scattering is taken into account in the coherent cross
section~(\ref{eq:xcoh_def}). This is a very good approach at
$\varepsilon\gtrsim{}1$~meV. For a perfect lattice, the coherent elastic
cross section per one molecule has the form of a~sum over the
reciprocal-lattice vectors $\vec{\tau}$ \cite{love84}
\begin{equation}
  \label{eq:xBragg_nb}
    \left(\frac{\mathrm{d}\sigma}{\mathrm{d}\Omega}\right)%
    _{\!\mathrm{coh}}^{\!\mathrm{el}} 
    = \frac{(2\pi)^3}{V_0} \sum_{\vec{\tau}}|F_N(\vec{\tau})|^2
    \delta(\vec{\kappa} - \vec{\tau}) \,
    \exp(-2W) \,,
\end{equation}
where $V_0$ denotes the unit-cell volume and $|F_N(\vec{\tau})|^2$ is
the average squared unit-cell factor
\begin{equation}
  \label{eq:cell_fac}
  |F_N(\vec{\tau})|^2 = \sigma_\mathrm{coh} \, \frac{1}{N}
  \Bigl|\sum_{d=1}^{N} \exp(i\vec{\tau}\cdot\vec{d})\Bigr|^2.
\end{equation}
Vector $\vec{d}$ denotes here the position of a given molecule in the
unit cell and $N$ is the number of molecules per unit cell. The unit
cell of hcp sD$_2$ lattice contains two molecules. From
Eq.~(\ref{eq:xBragg_nb}) it follows that the Bragg scattering takes
place only if the condition
\begin{equation}
  \label{eq:cond_Bragg_kappa}
  \vec{\kappa} = \vec{k} - \vec{k}' = \vec{\tau} 
\end{equation}
is fulfilled. This condition cannot possibly be satisfied below the
Bragg-cutoff energy
\begin{equation}
  \label{eq:e_Bragg-cutoff}
  \varepsilon_\mathrm{B} = \tau_\mathrm{min}^2/(8 m_n) \,,
\end{equation}%
in which $\tau_\mathrm{min}$ denotes the value of the shortest vector
$\vec{\tau}$ with $F_N(\vec{\tau})\neq{}0$. In the case of hcp
crystal with the following basic vectors of the inverse lattice
\begin{equation}
  \label{eq:basic_vect}
  \begin{split}
    \vec{\tau}_1 & = c\, (1,1/\sqrt{3},0), \qquad 
    \vec{\tau}_2 = c\, (0,2/\sqrt{3},0), \\
    \vec{\tau}_3 & = c\, (0,0,\sqrt{3/8}),
  \end{split}
\end{equation}
where $c = 2\pi/a_\mathrm{hcp}$, the Bragg-cutoff energy is determined
by the vector $\vec{\tau}_\mathrm{min}=[1,0,0]$. This corresponds to
$\varepsilon_\mathrm{B}\approx{}2$~meV for sD$_2$ at low pressure.

When sD$_2$ has a random polycrystalline structure, the cross section
(\ref{eq:xBragg_nb}), derived for a monocrystal sample, can be averaged
over all orientations of the lattice. This leads to the polycrystalline
cross section
\begin{equation}
  \label{eq:xBragg_poly}
    \left(\frac{\mathrm{d}\sigma}{\mathrm{d}\Omega}\right)%
    _{\!\mathrm{coh}}^{\!\mathrm{el}} = 
     \, \frac{2\pi^2}{V_0}\,\frac{1}{k^2}\,
    \sum_{\vec{\tau}} |F_N(\vec{\tau})|^2 \, \frac{1}{\tau} \\ 
    \delta\!\left(1-\frac{\tau^2}{2k^2}-\cos\vartheta\right)
    \exp(-2W) \,,
\end{equation}
$\vartheta$ being the angle between vectors $\vec{k}$ and $\vec{k}'$. If
a mosaic sD$_2$ monocrystal is considered, Eq.~(\ref{eq:xBragg_nb}) is
numerically averaged over the Gaussian distribution 
\begin{equation}
  \label{eq:mosaic_distr}
  g(\alpha) = \frac{1}{\alpha_0\sqrt{2\pi}}
  \exp\left(-\frac{\alpha^2}{2\alpha_0^2}\right) ,
\end{equation}
of angular deviation $\alpha$ of the mosaic blocks from the mean
direction. The standard deviation~$\alpha_0$, which is usually called
the mosaic spread, is here a~free parameter.

The total differential cross section for neutron scattering in sD$_2$ is
a sum of the incoherent cross sections (\ref{eq:x_rovib}) and
(\ref{eq:xinc_def}), with $\mathcal{S}_i$ given by
Eq.~(\ref{eq:Sinc_phon_exp}), and the coherent Bragg cross
section(\ref{eq:xBragg_nb}). For fixed initial energy $\varepsilon$ and
rotational state~$K$, all possible rotational and vibrational
transitions are included in this cross section. The total
cross section $\sigma_\mathrm{tot}$ is thus given as
\begin{equation}
  \label{eq:xsec_tot}
  \sigma_\mathrm{tot} = \int \mathrm{d}\varepsilon'\mathrm{d}\Omega \, 
  \left[ \sum_n 
    \left(\frac{\partial^2\sigma}
    {\partial\varOmega\partial\varepsilon'}\right)_{0n}
  +\left(\frac{\partial^2\sigma}
    {\partial\varOmega\partial\varepsilon'}\right)_{\!\mathrm{incoh}}
   \right] 
   + \int \mathrm{d}\Omega \,
  \left(\frac{\mathrm{d}\sigma}{\mathrm{d}\Omega}\right)%
  _{\!\mathrm{coh}}^{\!\mathrm{el}},
\end{equation}
where the coherent fraction is calculated using Eq.~(\ref{eq:xBragg_nb})
or Eq.~(\ref{eq:xBragg_poly}), depending on the structure of a sample.
If several rotational states are initially populated, the incoherent
fraction of $ \sigma_\mathrm{tot}$ is additionally averaged over a
distribution of these states. At
$\varepsilon\gg{}\varepsilon_\mathrm{B}$, the coherent effects are
negligible and therefore the total cross section is solely determined by
the incoherent processes. At $\varepsilon\gg\omega_\mathrm{D}$, effects
of molecular binding in the lattice can be neglected and thus the cross
section for sD$_2$ (per one molecule) corresponds to that for a free
D$_2$ molecule.

\section{Results}
\label{sec:results}

Below, the calculated cross sections are compared with the experimental
total cross sections for cold neutrons~\cite{kasp08,atch09}, which were
determined from neutron-beam attenuation in sD$_2$ samples. The purpose
of the experiment was to study sD$_2$ as the UCN converter. The
measurements were performed in almost pure ortho-D$_2$
($98.7\pm{}0.2$\%), since a significant admixture of para-D$_2$ would
drastically decrease the UCN lifetime~\cite{liu00}, due to the neutron
energy gain during the rotational relaxation $K=1\to{}K'=0$ in D$_2$.
Therefore, the present calculations have also been performed for pure
ortho-D$_2$. In comparison with the experimental data it is assumed that
the sD$_2$ target is small so that the neutron leaves the crystal after
a single scattering and thus the kinematic diffraction theory is
applicable. All available data are normalized to the value of 4.1~b at
$\lambda=6.5$~\AA\ and $T=8$~K~\cite{kasp_priv}, which is due to
problems in accurate normalization of the measured cross sections.

\begin{figure}[htb]
  \begin{center}
    \includegraphics[width=9cm]{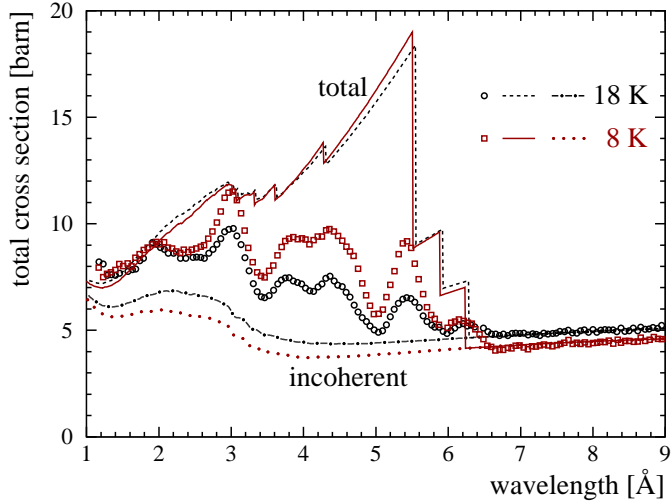}
    \caption{Calculated (solid and dashed lines) and experimental
      (squares and circles, from Fig.~4 of Ref.~\cite{atch09}) total
      cross sections for cold neutron scattering in a polycrystalline
      ortho-D$_2$ at temperature $T=8$~K and 18~K as functions of
      neutron wavelength. Also are shown the calculated incoherent
      contributions to the total cross sections (dotted and dot-dashed
      lines).
      \label{fig:politemp}}
  \end{center}
\end{figure}

The calculated total cross section for cold neutron scattering in a
polycrystalline sD$_2$ at temperature $T=8$~K and 15~K is shown in
Fig~\ref{fig:politemp}.
One can see that a very good agreement with experiment is reached for
$\lambda\gtrsim{}6.5$~\AA, where there is no Bragg scattering and the
total cross section is practically determined by the incoherent
processes. Also, a good agreement is visible for
$\lambda\gtrsim{}2$~\AA, where a contribution of the Bragg scattering to
the total cross section is not dominant. The incoherent cross section is
greater at $T=18$~K since the phonon processes are more probable at
higher temperatures. On the other hand, the calculated Bragg cross
section is greater at $T=8$~K because the Debye-Waller factor is smaller
at higher temperatures. Similar results were obtained in
Ref.~\cite{gran09}, with the use of $Z(\omega)$ derived from experiments.

\begin{figure}[htb]
  \begin{center}
    \includegraphics[width=9cm]{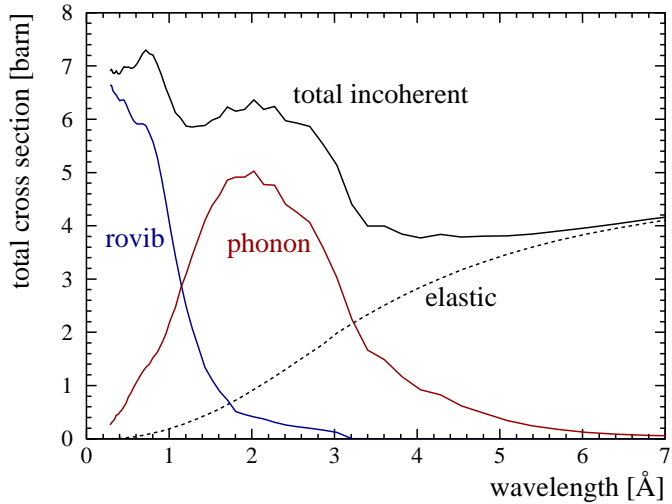}
    \caption{Calculated contributions to the total incoherent cross
      section as functions of the neutron wavelength, for a solid
      ortho-D$_2$ at $T=8$~K. The label ``elastic'' corresponds to the
      incoherent strictly-elastic scattering. A sum of the phonon
      creation and annihilation cross sections, with $\Delta{}E=0$, is
      denoted by the label ``phonon''. A sum of the
      rotational-vibrational transitions with possible simultaneous
      phonon processes is represented by the label ``rovib''.
      \label{fig:xinc_8k}}
  \end{center}
\end{figure}

Different contributions to the total incoherent cross section are shown
in Fig.~\ref{fig:xinc_8k}, for a solid ortho-D$_2$ at $T=8$~K.
The interval $\lambda\approx{}4$--7~\AA\ is dominated by the incoherent
strictly-elastic scattering. The cross section of this process is
described by the first term of expansion~(\ref{eq:Sinc_phon_exp}). The
purely phonon processes are most significant in the vicinity of
$\lambda=2$~\AA. The rotational-vibrational excitations begin to
dominate the incoherent cross section above the first rotational
threshold at $\lambda\approx{}1$~\AA, which corresponds to the
excitation $K=0\to{}K'=1$. Since in sD$_2$ this excitation may take
place with simultaneous phonon annihilation, the respective cross
section is visible up to $\lambda\approx{}3$~\AA.

The theoretical curves in Fig.~\ref{fig:politemp} suggest that the
experimental samples were not random sD$_2$ polycrystals. In the
interval $\lambda\approx{}2$--6.5~\AA, the experimental data display
several broad peaks with similar widths. The magnitudes of theoretical
sharp peaks are mostly much larger and practically do not depend on
temperature. Their locations do not fully coincide with the locations of
observed peaks. In order to understand this problem, it is reasonable to
consider neutron scattering in sD$_2$ monocrystal at various
orientations. In comparison of theory with experiment, it is necessary
to take into account the wavelength uncertainty
$\Delta\lambda\approx{}0.05$~\AA~\cite{kasp_priv}. This value is too
small to explain the presence of broad Bragg peaks in a monocrystal
sample. Therefore, a possible mosaic structure in the sD$_2$ monocrystal
is considered, which can lead to a significant broadening for a
sufficiently large value of the mosaic spread.

The calculated total cross section for cold neutron scattering from
\begin{figure}[htb]
  \begin{center}
    \includegraphics[width=9cm]{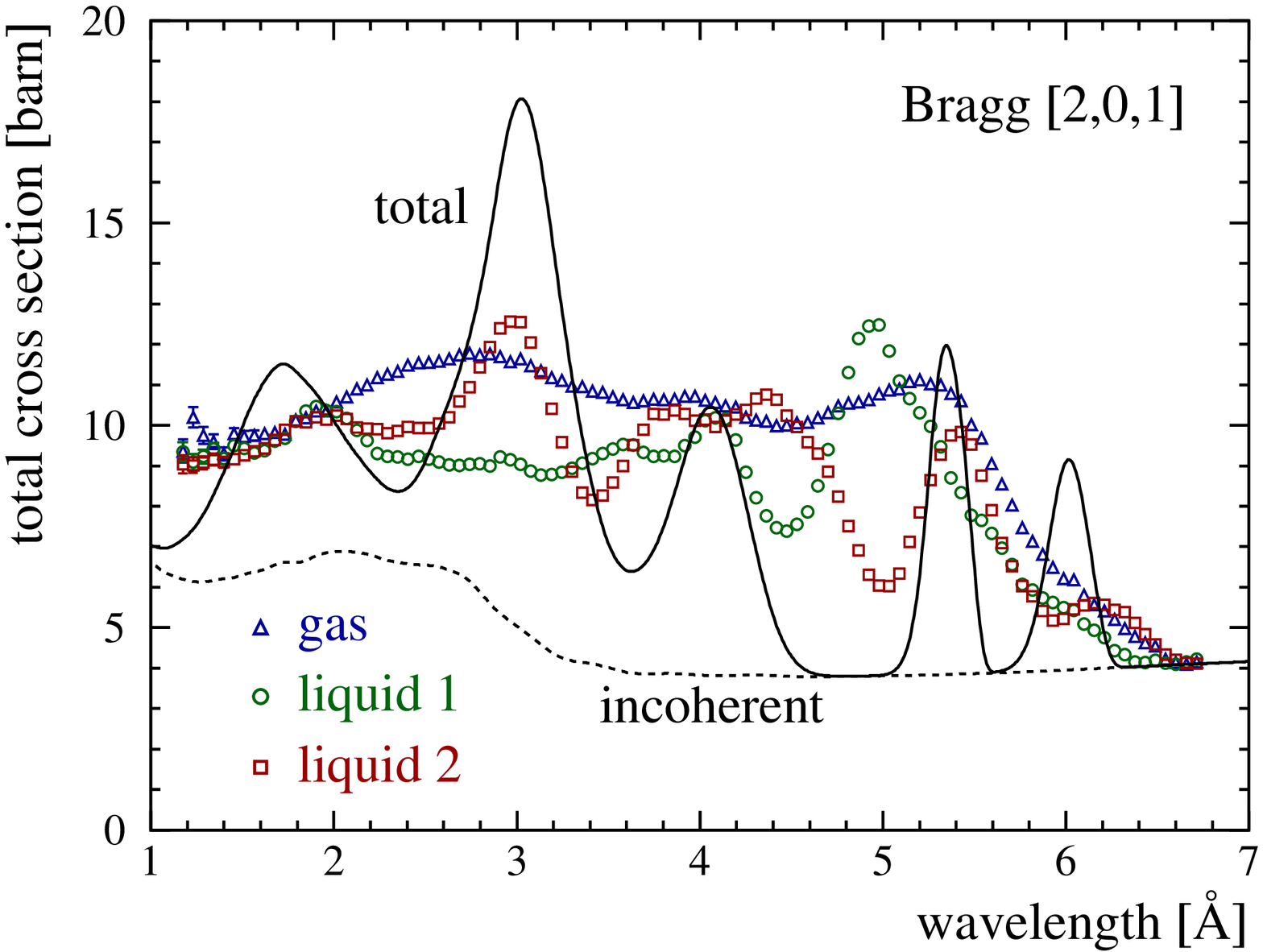}
    \caption{Calculated total and incoherent cross section for cold
      neutron scattering from a 8-K ortho-D$_2$ monocrystal, oriented in
      the [2,0,1] direction, as a function of neutron wavelength. The
      mosaic spread $\alpha_0=3^{\circ}$ has been assumed. The
      experimental data for $T=8$~K (from Fig.~5 of Ref.~\cite{atch09})
      were determined using three different sD$_2$ targets, obtained
      by freezing D$_2$ gas (one sample) and D$_2$ liquid (two
      samples).
      \label{fig:xexpt_201_8k}}
  \end{center}
\end{figure}
a~8-K sD$_2$ monocrystal oriented in the [2,0,1] direction is shown in
Fig.~\ref{fig:xexpt_201_8k}. A mosaic structure of the monocrystal,
with the spread $\alpha_0=3^{\circ}$, has been assumed. Also, the
wavelength resolution is taken into account. The plotted experimental
data~\cite{kasp08,atch09} for the three 8-K samples demonstrate that the
specific structures of sD$_2$ were different, although freezing
conditions were very similar. The two samples were frozen from liquid
D$_2$ (labels ``liquid~1'' and ``liquid~2'') and one sample was obtained
by freezing D$_2$ gas (label ``gas''). The presence of various patterns
of Bragg scattering is obvious, if the considered samples were, at least
in part, monocrystals grown in different orientations with respect to
the incident neutron beam. Some peaks of the calculated cross section
approximately correspond to the peaks observed in the ``liquid~2''
target. On the other hand, the calculated Bragg maximum at 4~\AA\ is
visible in the ``liquid~1'' case. The mosaic spread of about~$3^{\circ}$
is optimal --- an increase or decrease by~$1^{\circ}$ leads to a worse
agreement with the experiment. As one may expect, in the limit
$\alpha_0\to{}180^{\circ}$ the theoretical cross section tends to the
polycrystalline pattern, independently of a chosen crystal orientation.

The cross sections calculated for all the orientations $[h,k,l]$ with
$h$, $k$, $l=0,1,\ldots,10$ do not accurately reveal the observed Bragg
patterns of the three samples. This suggest that these samples could be
composed of a~certain number of large mosaic monocrystals. Visible
cracks and optical opacity of the sD$_2$ targets~\cite{kasp08,bode04},
which vary with temperature, support such a~model.
\begin{figure}[htb]
  \begin{center}
    \includegraphics[width=9cm]{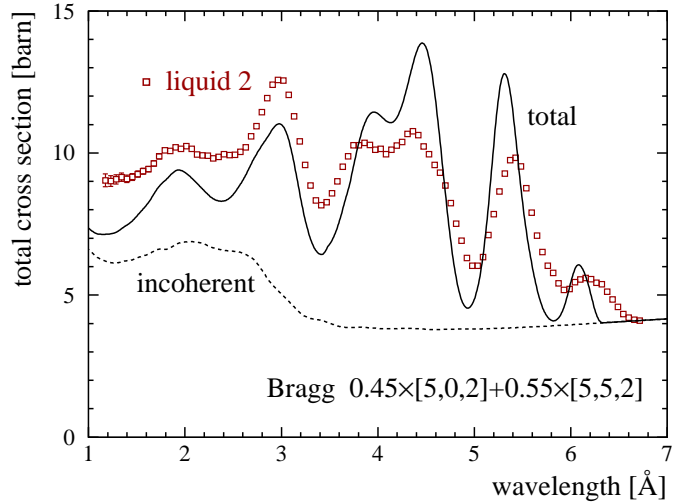}
    \caption{Calculated total and incoherent cross section for cold
      neutron scattering from a 8-K ortho-D$_2$ target, which is
      composed of large hcp monocrystals oriented in the directions
      [5,0,2] (45\%) and [5,5,2] (55\%). The mosaic spread
      of~$3^{\circ}$ is assumed. The experimental 8-K data were measured
      using the sD$_2$ sample grown from liquid D$_2$ (``liquid~2'',
      Fig.~5 from Ref~\cite{atch09}).
    \label{fig:xsum_liq2}}
  \end{center}
\end{figure}
The total cross section in Fig.~\ref{fig:xsum_liq2} has been obtained
for the superposition of two crystal orientations: [5,0,2] with the
probability $\mathcal{P}=0.45$ and [5,5,2] with $\mathcal{P}=0.55$. The
theory quite well reveals the locations of Bragg peaks for the
``liquid~2'' sample. The differences between magnitudes of the
calculated and experimental peaks do not exceed 30\%.  
\begin{figure}[htb]
  \begin{center}
    \includegraphics[width=9cm]{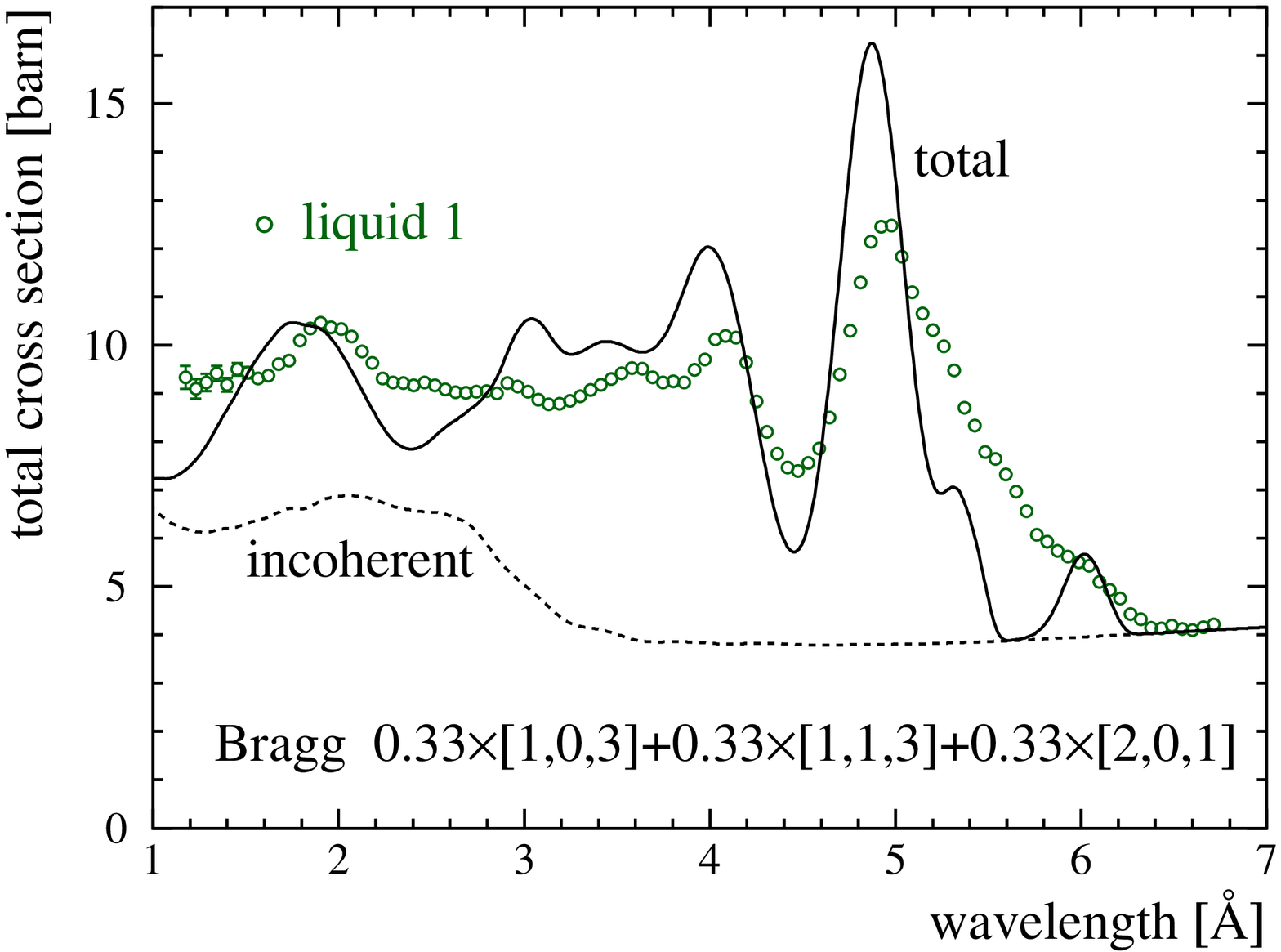}
    \caption{Calculated total and incoherent cross section for cold
      neutron scattering from a 8-K ortho-D$_2$ target, which is
      composed of large hcp monocrystals oriented in the directions
      [1,0,3] (33\%), [1,1,3] (33\%), and [2,0,1] (33\%).  The mosaic
      spread of~$3^{\circ}$ is assumed. The experimental 8-K data were
      measured using the sD$_2$ sample grown from liquid D$_2$
      (``liquid~1'', Fig.~5 from Ref~\cite{atch09}).
      \label{fig:xsum_liq1}}
  \end{center}
\end{figure}
In the ``liquid~1'' case, which is presented in
Fig.~\ref{fig:xsum_liq1}, an approximate description of the data has
been achieved assuming the presence of three following monocrystal
orientations in the sample: [1,0,3], [1,1,3], and [2,0,1]. Here they
have the same probability $\mathcal{P}=0.33$.

The peaks of the measured cross section for the ``gas'' sample are less
pronounced than the peaks observed in the ``liquid~1'' and ``liquid~2''
targets. This can probably be ascribed to the lack of local correlations
between D$_2$ molecules in gaseous deuterium. As a result, much more
various orientations of sD$_2$ monocrystals are present in the ``gas''
sample, compared to the targets grown from liquid deuterium.
\begin{figure}[htb]
  \begin{center}
    \includegraphics[width=9cm]{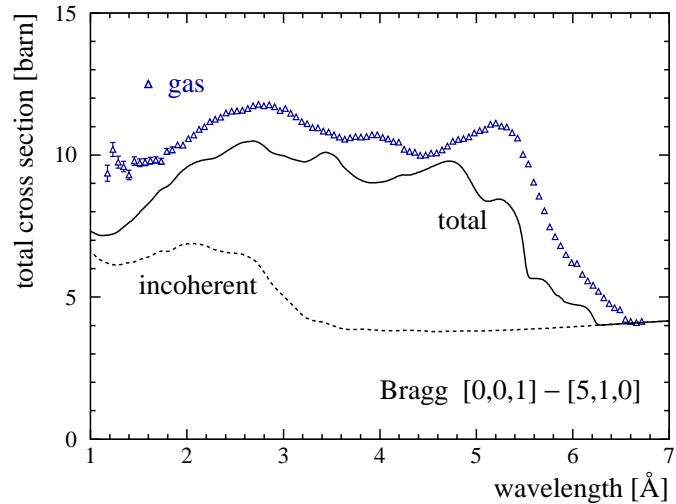}
    \caption{Calculated total and incoherent cross section for cold
      neutron scattering from a 8-K ortho-D$_2$ target, which is
      composed of large hcp monocrystals oriented in all directions
      corresponding to the 69 shortest vectors of the inverse lattice
      (in the range [0,0,1]--[5,1,0]).  The mosaic spread of~$3^{\circ}$
      is assumed. The experimental 8-K data were measured using the
      sD$_2$ sample grown from a~gaseous D$_2$ (``gas'', Fig.~5 from
      Ref~\cite{atch09}).
      \label{fig:xsum_gas}}
  \end{center}
\end{figure}
The theoretical cross section in Fig.~\ref{fig:xsum_gas} has been
calculated assuming that the ``gas'' target is composed of the
monocrystals that have orientations corresponding to the 69 shortest
vectors of the inverse lattice (in the range [0,0,1]--[5,1,0]).  It is
assumed that all the orientations have the same probability. The mosaic
spread of~$3^{\circ}$ in every single monocrystal is also taken into
account. One can see that the calculated total cross section roughly
corresponds to the ``gas'' data. A better agreement with the experiment
can be achieved by tedious adjusting the probabilities of various
orientations, which are free parameters, and by using more monocrystal
orientations, especially in the ``liquid~1'' and ``liquid~2'' cases.
However, this is not a purpose of this work.

The experimental data shown in Fig.~\ref{fig:politemp} were measured
using the same ``liquid~2'' sample at $T=8$~K and 18~K. The Bragg
patterns for different temperatures, including 14~K and
17~K~\cite{kasp08,atch09}, are very similar. This means that the mosaic
spread and orientations of macroscopic monocrystals do not significantly
change, for a fixed sample. On the other hand, the magnitudes of
observed Bragg peaks systematically decrease with rising temperature.
This decrease is appreciable, especially when compared to the incoherent
cross section which increases with rising temperature. The incoherent
processes are well described by the theory. However, the calculated
Bragg cross sections do not reveal strong temperature changes of the
measured cross sections, both for the random-polycrystalline and
monocrystal cases. A~possible explanation is that multiple Bragg
scattering took place in the sD$_2$ targets and, thus, the kinematic
diffraction theory is not fully applicable to the considered experiment.
In such cases, phenomena of primary and secondary extinction, which
strongly depend on crystal structure and change the neutron-beam
attenuation, should be taken into account (see e.g., Ref.~\cite{sear77}
and references therein).
\begin{figure}[htb]
  \begin{center}
    \includegraphics[width=9cm]{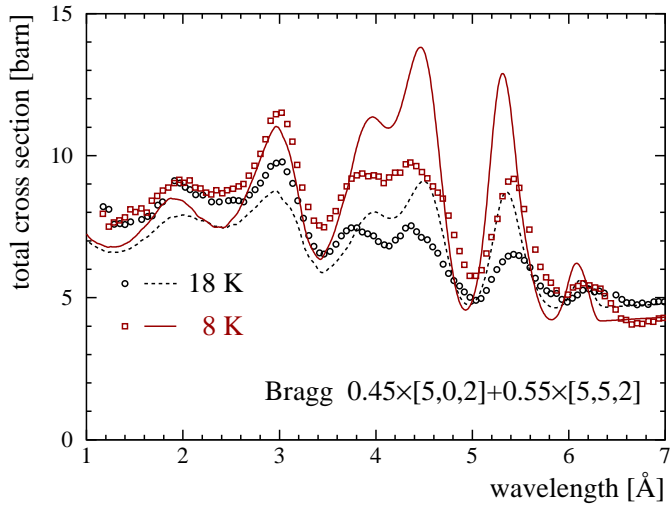}
    \caption{The same as in Fig.~\ref{fig:xsum_liq2} for
      $T=8$ and 18~K of the ``liquid~2'' sample.  A~contribution from
      the Bragg scattering to the calculated total cross section at
      $T=18$~K is scaled by the factor of~0.5.
      \label{fig:monotemp}}
  \end{center}
\end{figure}
In Fig.~\ref{fig:monotemp}, the calculated total cross sections for the
``liquid~2'' sample at 8~K and 18~K are compared with the experimental
cross sections. A qualitative agreement of theory and experiment is
achieved here by scaling the 8-K Bragg cross section by a factor of~0.5.

\section{Conclusions}

The calculated cross sections for cold neutron scattering in solid
deuterium are in a reasonable agreement with the experimental data, if
it is assumed that the sD$_2$ targets are composed of limited numbers of
mosaic monocrystals with the mosaic spread of about~$3^{\circ}$. In
particular, this naturally explains the presence of different Bragg
patterns in the sD$_2$ samples which are grown separately, although
under very similar conditions. Such patterns are due to various
orientations of the monocrystals, with respect to the impinging neutron
beam. The magnitude of incoherent cross section and its variation with
temperature is well described by the theory. On the other hand, the
magnitude of measured coherent elastic cross section decreases strongly
with rising temperature, which is not revealed by the theory.
A~possible explanation of this disagreement is that the kinematic
diffusion theory is not valid for comparison of the theoretical cross
sections with the experimental cross sections, which are determined
using the neutron-beam attenuation method.

Although the presented cross sections have been calculated in the
Bragg-scattering region, the conclusions regarding the sD$_2$ structure
are of importance for scattering of ultracold neutrons and thus for
developing high-intensity UCN sources. The phonon annihilation in
ortho-D$_2$ is a~crucial process which leads to UCN upscattering and,
therefore, limits the effectiveness of sD$_2$ converters. Since the
amplitude of neutron scattering from deuterium is mostly coherent, it is
necessary to take into account a real target structure in the UCN-energy
region. In estimations of the coherent phonon annihilation, the
polycrystalline approach and neglecting crystal imperfections may lead
to strong disagreement between theory and experiment. Figure~2 of
Ref.~\cite{atch05a}, which shows strong changes of the UCN cross
sections for different 5-K sD$_2$ samples, is an example of such
disagreement.


\section*{Acknowledgments}
I am indebted to M.~Kasprzak for supporting the experimental data and
explanations. K.~Bodek and A.~Kozela are acknowledged for discussions.



\end{document}